\documentclass{article}

\usepackage[english]{babel}

\usepackage[letterpaper,top=2cm,bottom=2cm,left=3cm,right=3cm,marginparwidth=1.75cm]{geometry}

\usepackage{amsmath,amsfonts,amssymb}
\usepackage{graphicx}
\usepackage[colorlinks=true, allcolors=blue]{hyperref}
\usepackage{algorithm, algpseudocode}

\usepackage{tikz}
\usetikzlibrary{quantikz2}
\usepackage{adjustbox}
\usepackage{authblk}

\newtheorem{definition}{Definition}[section]

\title{Benchmarking Techniques for Decoded Quantum Interferometry}
\author[1]{Leon Bollmann}
\author[1]{Maximilian Hess}
\affil[1]{Infineon Technologies AG, Neubiberg, Germany}
\date{}

\begin{document}
\maketitle

\begin{abstract}
We develop a new benchmarking scheme for the Decoded Quantum Interferometry (DQI) algorithm quantifying the number of quantum gates required to obtain an optimal solution to a problem amenable to DQI. We apply the benchmarking scheme to the Binary Paint Shop Problem (BPSP) in order to benchmark the performance of DQI against a state of the art classical solver. To do so, we provide an explicit construction of a quantum circuit implementation of a greedy decoder for low-density parity check codes arising from max-2-XORSAT problems.
\end{abstract}

\section{Introduction}

For a long time, quantum algorithms for combinatorial optimization were largely based on one of two principles.
The first being variational quantum algorithms~\cite{Cerezo_2021}, where a parameterized quantum circuit is executed multiple times with the goal of optimizing the expectation of an observable in correspondence with an optimization problem. The most prominent variational optimization algorithm being the Quantum Approximate Optimization Algorithm (QAOA)~\cite{farhi2014quantumapproximateoptimizationalgorithm}.
The second approach is speeding up (parts of) classical optimization algorithms with quantum search algorithms based on Grover's search algorithm~\cite{Grover1996FastQuantumSearch} or Amplitude Amplification~\cite{Brassard_2002}.
A general framework for quantum search in optimization is provided in~\cite{Gilliam_2021} while quantum versions of more elaborate classical schemes have been suggested in~\cite{Montanaro2018QuantumWalkBacktracking, Montanaro_2020}.

An entirely different approach, called Decoded Quantum Interferometry (DQI), has recently been suggested in~\cite{Jordan_2025}.
Given a prime $p$, a natural number $n\in\mathbb{N}$ and an objective function $f: \mathbb{F}_p^n \rightarrow \mathbb{Z}$ of a certain structure, DQI produces a state proportional to
\begin{equation}\label{eq:dqi_state}
    \sum_{x \in \mathbb{F}_p^n} P(f(x)) \ket{x}.
\end{equation}
The amplitudes of a state are modified using a polynomial $P$ of specifiable degree.
This can be used to ``boost" the measurement probabilities of states with high or low objective values.
The most general setting for which this procedure is shown to work is max-LINSAT, a constraint satisfaction problem for linear constraints over the finite field $\mathbb{F}_p$.
A special case of max-LINSAT which exhibits provable super-polynomial speedups~\cite{Jordan_2025} is the Optimal Polynomial Intersection (OPI) problem.
This work will focus on another special case, namely those instances of max-LINSAT where $p=2$, also called max-XORSAT.
The main novelty behind DQI is that it reduces an optimization problem to a decoding problem. The preparation of the target state~\eqref{eq:dqi_state} is relatively straightforward except for one crucial step. First a weighted Dicke state is prepared using well known quantum circuits~\cite{Bartschi_2022}.
Then the problem data is weaved into the Dicke state by a series of well understood operations such as matrix multiplication. Then follows the crucial ``decoding" step where errors $y \in \mathbb{F}_2^m$ have to be uncomputed using the syndrome $B^T y \in \mathbb{F}_2^n$. As in general the number of constraints $m$ can be larger than the number of variables $n$, this is not always possible without errors.
From the decoded quantum state, a Hadamard transform yields the DQI target state~\eqref{eq:dqi_state}.
A more detailed review of DQI is given in section~\ref{subsec:dqi_recap}.

Subsequent works have explored practical and theoretical directions in connection with DQI.
For the OPI problem examined in the original DQI proposal, ~\cite{khattar2025verifiablequantumadvantageoptimized} provides sophisticated decoding circuits and demonstrate verifiable speedup over classical solvers.
The implementation of costly steps in the DQI protocol is further simplified to yield a more efficient algorithm in~\cite{rosmanis2026nearlylineartimedecodedquantum}.
On the other hand, limitations of DQI have been explored, via efficient classical counterparts~\cite{Parekh_2025} and proof of the inability of polynomial-time algorithms to outperform random guessing on max-LINSAT~\cite{kramer2026tightinapproximabilitymaxlinsatimplications}.

On a more practical note, in~\cite{sabater2025solvingindustrialintegerlinear}, the authors provide a general framework for embedding any Integer Linear Program (ILP) into the max-XORSAT format, implement a general purpose Belief Propagation Decoder as a quantum circuit and provide benchmarking for the automotive bundling problem.

The main contribution of this work consists in a new benchmarking method for DQI which focuses on determining amplitudes
of classically chosen states (e.g. optimal solutions) instead of ``average" performance measures like the expected number of satisfied clauses, as provided by the original paper~\cite{Jordan_2025}. 
We argue that determining the probability to measure an optimal or near-optimal solution is closer to what optimization practitioners are actually interested in and that expectation values for the solution quality hide a lot of relevant information.
Our benchmarking scheme is similar to the one suggested for Grover-style algorithms in~\cite{Cade_2023} as DQI, just like Grover's algorithm, allows us to track amplitudes of select states through the algorithm without having to simulate (or physically execute) the whole quantum protocol.

For small problems, it is possible to exactly determine the measurement probability of a given bit string, such as the optimal solution and the polynomial degree $l$. 
In the original paper, the authors provide an efficient classical method of determining the degree-$l$ polynomial which boosts the good solutions in an optimal way~\cite{Jordan_2025}[Lem.~9.2]. This is provably the case for the case of perfect decoding, and we choose to stick to the same method in the practically relevant case of imperfect decoding.
We show how the choice of Dicke state weights $w_0,\dots,w_l$ which is explicitly given in the original source leads to the coefficients of the polynomial $P$.
Using our explicit expression for $P$ we can calculate the amplitudes and measurement probabilities for optimal solutions $x \in \mathbb{F}_2^n$ in a straightforward manner, namely by evaluating $P(f(x))$.
We provide a concrete description of this method in section~\ref{subsec:exact_benchmarking}.

An exact computation of the coefficients of degree-$l$ polynomial $P$ requires us to explicitly determine the errors of Hamming weight $k \leq l$ which are decoded incorrectly by the quantum decoding method of our choice.
Following the ``semicircle law" in~\cite{Jordan_2025} we expect that the polynomial degree $l$ must grow linearly with the problem size.
Evaluating the decoder for $\sum_{k=0}^l \binom{n}{k} \in \Theta(n^l)$ errors therefore becomes prohibitive for larger problems.
We propose an alternative, approximate method of still obtaining relevant benchmarks in section~\ref{subsec:approximate_benchmarking}.
It is based on performing a Monte-Carlo approximation for the error rate by evaluating the decoder on a constant number of errors for each Hamming weight.
As the formulae for the decoder coefficients additionally require knowing which errors are decoded incorrectly, we further assume that they are uniformly distributed in each Hamming shell. We are then able to determine the polynomial coefficients and therefore the measurement probabilities in expectation with respect to the assumed uniform distribution.
Figure~\ref{Fig_Approximation} shows that this leads to a slightly conservative but relatively accurate approximation to the actual measurement probabilities obtained for smaller instances.

We proceed to showcase our benchmarking methods using the Binary Paint Shop Problem (BPSP) as an example.
A hypothetical automotive paint shop can color cars in two colors, say blue ($0$) and red ($1$).
Given $n$ cars, each of which appears twice in an arbitrary but fixed order, the task is to color each car once in blue and once in red while minimizing the number of color changes.
The BPSP was chosen as it admits a relatively straightforward max-XORSAT formulation.
We introduce the problem in detail and examine two different encodings for the BPSP and the resulting codes in section~\ref{sec_BPSP}.
We find that both encodings lead almost certainly to a constant code distance, i.e. minimal number of linearly dependent constraints.
The code distance is an upper bound for the Hamming weight up to which errors can be decoded correctly in principle.
This finding makes it clear that it is necessary to explore the imperfect decoding regime.

We discuss decoding strategies for codes resulting from the BPSP in section~\ref{subsec:bpsp_decoder}. In either problem encoding, each constraint involves exactly $2$ variables which enables us to translate the decoding problem into a graph structure $G$ where the variables and constraints are represented by vertices and edges, respectively.
The decoding problem then boils down to finding a so called $T$-join of $G$ where $T$ is the set of indices for which the syndrome $B^T y$ has the value $1$.
This problem can be further reduced to a minimum weight matching problem on a modified version of $G$.
The minimum weight matching problem can be solved in polynomial time by the Blossom Algorithm~\cite{Edmonds_1965, Edmonds_1965_weighted}.
However, we chose to examine a simpler, greedy method for the minimum weight matching problem as it allows for a significant amount of information to be efficiently precomputed on a classical computer.
The remaining algorithm can be realized relatively straightforwardly as a quantum circuit using $\mathcal{O}(n^2)$ auxiliary qubits.
While the greedy matching algorithm does not optimally solve the problem in general, it runs almost quadratically faster than the Blossom Algorithm.

In section~\ref{sec:results}, we benchmark DQI for the BPSP using both decoders and find that the greedy version leads to higher probabilities of measuring the optimal solution.
This is due to the faster runtime of the greedy decoder which allows us to repeat the greedy DQI protocol more often within the same budget of quantum gates.
It should be noted that while we obtained benchmarking results for a Blossom decoder, these results remain hypothetical as we did not attempt to devise a quantum version of the relatively complex Blossom Algorithm.

Further, we compare different values for the polynomial degree $l$ and find that the optimal choice scales linearly with the problem size, in spite of deteriorating decoder performance for larger instances.
Finally, we compare the scaling of DQI for instances of up to $80$ pairs of cars with the scaling of the leading classical solver Gurobi~\cite{gurobi}. We find that Gurobi exhibits a significantly flatter scaling than either version of DQI.

\section{DQI benchmarking techniques}\label{sec:benchmarking}

\subsection{Summary of DQI algorithm}\label{subsec:dqi_recap}
We focus on the DQI algorithm for max-XORSAT problems and denote by $\mathbb{F}_2$ the finite field with two elements. Given a constraint matrix $B\in\mathbb{F}_2^{(m\times n)}$ and a vector $v\in\mathbb{F}_2^m$, the max-XORSAT problem is to find an assignment of variables $x\in\mathbb{F}_2^n$ which satisfies as many of the $m$ constraints corresponding to $B$ and $v$ as possible. This can also be written as $x$ maximizing the objective function
\begin{equation}
    f(x):=\sum_{j=1}^m f_j(b_j\cdot x):=\sum_{j=1}^m(-1)^{v_j+b_j\cdot x},
\end{equation}
where $b_j$ denotes the $j$th row of the matrix $B$. In order to solve this problem, the DQI algorithm aims to prepare a state 
\begin{equation}
    \ket{P(f)} := \sum_{x\in\mathbb{F}_2^n}P(f(x))\ket{x},
\end{equation}
which boosts the probability of measuring good solutions to the given max-XORSAT problem via an application of a degree $l$ polynomial $P$ with action $P(\cdot):=\sum_{k=0} ^l\alpha_k (\cdot)^k$. A crucial step in the preparation of the above state can be thought of as syndrome decoding with respect to the low-density parity-check (LDPC) code generated by the parity matrix $B^\intercal$. As this decoding step fails to succeed without errors for general max-XORSAT instances, in particular for the BPSP studied in the present article, we obtain an imperfect DQI state $\ket{P_D(f)}$, instead of the state $\ket{P(f)}$. Still, this imperfect DQI state enhances the probability of measuring a good or even optimal solution to the given max-XORSAT problem. We now give a brief overview of the required steps to prepare the state $\ket{P_D(f)}$.
\begin{enumerate}
\item Classical precomputation of coefficients $w_k$: \\
The coefficients $\alpha_k$ of the polynomial $P$ are do not enter directly in the algorithm. Instead coefficients $w_k$, $0\leq k \leq l$, are computed which relate to the coefficients $\alpha_k$ through the decomposition of $P$ into elementary symmetric polynomials. As shown in \cite{Jordan_2025}[Thm.~4.1] there is an optimal choice of $w_k$ under the assumption of perfect decoding. In the case of imperfect decoding one can still choose the coefficients $w_k$ in the same way even though this is no longer optimal. This is done in \cite{Jordan_2025} and we follow this approach here. The coefficient vector $w=(w_0,\dots,w_l)^\intercal$ is then chosen to be the principal eigenvector of the symmetric tridiagonal $(l+1)\times (l+1)$ matrix
\begin{equation}
    A^{(m,l)} = \left(\begin{array}{cccccc}
       0  & a_1 & 0 & 0 & \dots & 0 \\
       a_1  & 0 & a_2 & 0 &\dots & 0 \\
       0 & a_2 & 0 & a_3 & \dots & 0 \\
       0 & 0 & a_3 & 0 & \dots & 0 \\
       \vdots & \vdots & \vdots & \vdots & \ddots & a_l \\
       0 & 0 & 0 & \dots & a_l & 0
    \end{array}\right)
\end{equation}
with $a_k =\sqrt{k(m-k+1)}$, $0\leq k \leq l$, see~\cite{Jordan_2025}[Lem.~9.2]. The coefficients can therefore be precomputed classically before preparing the DQI state.

\item Dicke state superposition, phase introduction and syndrome register: \\
We begin by preparing the state 
\begin{equation}
    \sum_{k=0}^l \frac{w_k}{\sqrt{\binom{m}{k}}} \sum_{y\in\mathbb{F}_2^m, \, |y|=k} (-1)^{v\cdot y}\ket{y}\ket{B^\intercal y},
\end{equation}
where $\binom{m}{k}=\frac{m!}{(m-k)!k!}$ denotes the binomial coefficient and $|y|:=\sum_{j=1}^m y_j$ denotes the Hamming weight.
To do so, we prepare the following weighted superposition of Dicke states
\begin{equation}
    \sum_{k=0}^l \frac{w_k}{\sqrt{\binom{m}{k}}} \sum_{y\in\mathbb{F}_2^m, \, |y|=k}\ket{y}.
\end{equation}
This step requires $O(m^2)$ quantum gates for $l \in \Theta(m)$. The main contributor to the cost is the circuit for efficient Dicke state preparation from~\cite{Bartschi_2022}.
A more detailed description of the state preparation can be found in~\cite{Jordan_2025}[sec.~8.1.2].

Next, we introduce a phase corresponding to the target vector $v$ and arrive at the state
\begin{equation}
    \sum_{k=0}^l \frac{w_k}{\sqrt{\binom{m}{k}}} \sum_{y\in\mathbb{F}_2^m, \, |y|=k} (-1)^{v\cdot y}\ket{y}
\end{equation}
Lastly, we compute the syndrome $B^\intercal y$ into an auxiliary register. This matrix-vector multiplication takes $O(nm)$ operations.

\item Decode Syndrome and uncompute $\ket{y}$ register: \\
This is the crucial step in the algorithm. The decoding step is the task of inferring $\ket{y}$ from $\ket{B^\intercal y}$. As $m>n$ in general, the mapping $y\mapsto B^\intercal y$ is in general not injective. Therefore, for a sufficiently high degree $l$ of the polynomial (depending on the quality of the associated LDPC code), there are bound to be errors $y\in\mathbb{F}_2^m$ of Hamming weight $|y|\leq l$ where the decoder fails to succeed. Sticking to the terminology of \cite{Jordan_2025}, we divide the $k$-Hamming shell $\mathcal{E}_k := \{y\in\mathbb{F}_2^m: \; |y|=k \}$, $k\in\mathbb{N}\cup \{0\}$, into disjoint subsets $\mathcal{D}_k:= \{y\in \mathcal{E}_k: \, y \ \text{decoded correctly}\}$ and $\mathcal{F}_k:= \{y\in \mathcal{E}_k: \, y \ \text{not decoded correctly}\}$. We also define the failure rate of the decoder on errors of Hamming weight $k$ as
\begin{equation}
    \epsilon_k := \frac{|\mathcal{F}_k|}{|\mathcal{E}_k|}.
\end{equation} After post-selection, we continue when the decoder was able to decode the error correctly. Therefore, we manage to successfully uncompute the error $\ket{y}$ if and only if $y\in\mathcal{D}_k$ for any $k\in\mathbb{N}\cup \{0\}$, and we are left with the state
\begin{equation}
    \frac{1}{\sqrt{R}}\sum_{k=0}^lw_k \sum_{y\in\mathcal{D}_k} (-1)^{v\cdot y} \ket{B^\intercal y},
\end{equation}
where $R:=\sum_{k=0}^l w_k^2(1-\epsilon_k)$ is the renormalization factor introduced through post-selection, see \cite{Jordan_2025}[Lem. 10.4]. The number of quantum gates required for this step depends on the choice of decoding algorithm. In our case, the leading order contribution of the quantum gates required for our decoder implementation will grow slightly faster than $m^2$ and hence also constitutes the leading-order contribution to the number of quantum gates required for the entire DQI algorithm. The details of the decoder implementation are contained in Section \ref{sec_BPSP}.

\item Hadamard transform: \\
Lastly, we apply the Hadamard transform to obtain the state
\begin{equation}
   \ket{P_D(f)} =  \prod_{j=1}^n H_j \frac{1}{\sqrt{R}}\sum_{k=0}^lw_k \sum_{y\in\mathcal{D}_k} (-1)^{v\cdot y} \ket{B^\intercal y},
\end{equation}
where $H_j$ denotes the Hadamard transform on the $j$th Qubit. We will refer to the state $\ket{P_D(f)}$ as the (imperfect) DQI state.
\end{enumerate}
\subsection{Exact benchmarking}\label{subsec:exact_benchmarking}


In \cite{Jordan_2025} asymptotic results are obtained for the expected number of constraints satisfied upon measuring the DQI state obtained after (imperfect) decoding. For the purposes of this work, we are however interested in a different type of benchmarking, focusing on the probability of obtaining an optimal solution to the problem upon measuring the DQI state. 
This is reminiscent of benchmarking techniques for Grover-style algorithms~\cite{Cade_2023} as we efficiently track the amplitudes of a small number of states throughout the algorithm without having to simulate the full DQI state at any point.
The measurement probabilities for specific states cannot be inferred from the expected number of satisfied constraints, hence the analysis in the present article does not utilize \cite{Jordan_2025}[Thm. 4.1, Thm. 10.1] but instead makes use of the following strategy. 
We only develop the benchmarking strategy for max-XORSAT problems, as we formulate the BPSP as a max-XORSAT problem. However, it would be straightforward to extend the benchmarking to the max-LINSAT formulation of DQI.

In case of perfect decoding, the DQI state $\ket{P(f)}$ is given by
\begin{equation}\label{DQI_state_perfect}
\ket{P(f)} = \sum_{x\in\mathbb{F}_2^n}\sum_{k=0}^l \frac{w_k}{\sqrt{2^n\binom{m}{k}}}\sum_{y\in\mathbb{F}_2^m, \,|y|=k} (-1)^{v\cdot y}(-1)^{(B^\intercal y) \cdot x}\ket{x},
\end{equation}
see \cite{Jordan_2025}[Sec. 8.1.1]. In case of imperfect decoding, only those errors $y\in\mathbb{F}_2^m$, $|y|\leq l$, which are decoded correctly contribute to the innermost sum in \eqref{DQI_state_perfect} after post-selecting on $y=\ket{0}$. We recall that we divide the Hamming shells $\mathcal{E}_k = \{y\in\mathbb{F}_2^m: \; |y|=k \}$, $k\in\mathbb{N}\cup \{0\}$ into disjoint subsets $\mathcal{D}_k= \{y\in \mathcal{E}_k: \, y \ \text{decoded correctly}\}$ and $\mathcal{F}_k= \{y\in \mathcal{E}_k: \, y \ \text{not decoded correctly}\}$. In a similar manner as for \eqref{DQI_state_perfect}, one sees that the resulting renormalized imperfect DQI state $\ket{P_D(f)}$ is then given as
\begin{equation}\label{DQI_state_imperfect}
\ket{P_D(f)} = \frac{1}{\sqrt{R}}\sum_{x\in\mathbb{F}_2^n}\sum_{k=0}^l \frac{w_k}{\sqrt{2^n\binom{m}{k}}}\sum_{y\in\mathcal{D}_k} (-1)^{v\cdot y}(-1)^{(B^\intercal y) \cdot x}\ket{x},
\end{equation}
where we recall $R=\sum_{k=0}^l w_k^2(1-\epsilon_k)$.
As the state $\ket{P_D(f)}$ is already given in the computational basis, we infer that the probability density $p_D(x)$ of a given assignment of variables $x\in \mathbb{F}_2^n$ of this state is given as
\begin{equation}\label{probability_assignment}
    p_D(x)=\frac{1}{R}\frac{1}{2^n}\sum_{k=0}^l \frac{w_k^2}{\binom{m}{k}}\bigg(\sum_{y\in\mathcal{D}_k} (-1)^{v\cdot y}(-1)^{(B^\intercal y) \cdot x}\bigg)^2.
\end{equation}
We are interested in the probability of finding an optimal solution to a given max-XORSAT problem. To evaluate this classically, we first compute the set of optimal solutions $S_{Opt}$ with a classical solver. The probability of obtaining any of these optimal solutions $p_{Opt}$ when measuring the imperfect DQI state is then simply given as
\begin{equation}
    p_{Opt} = \sum_{x\in S_{Opt}}p_D(x).
\end{equation}
We note that $p_D(x)$ is not constant on $S_{Opt}$ due to imperfect decoding. Although all $x\in S_{Opt}$ share the same objective value and therefore $\sum_{y\in\mathbb{F}_2^m, \,|y|=k} (-1)^{v\cdot y}(-1)^{(B^\intercal y) \cdot x}$ is constant on $S_{Opt}$, the corresponding sum $\sum_{y\in\mathcal{D}_k} (-1)^{v\cdot y}(-1)^{(B^\intercal y) \cdot x}$ is no longer constant on $S_{Opt}$.
In order to benchmark DQI against other solvers for the given max-XORSAT problem, we want to compare the expected number of operations it takes DQI to find an optimal solution to the problem against other solvers. Given that we have already computed $p_{Opt}$, we infer the expected number of times $c_{Opt}$ we need to run DQI in order to obtain an optimal solution by computing
\begin{equation}
    c_{Opt}=\sum_{j=1}^\infty j\cdot p_{Opt}\cdot(1-p_{Opt})^{j-1}=\frac{1}{p_{Opt}}.
\end{equation}
We further analyze the runtime $c_{DQI}$ of a single iteration of DQI by counting the number of quantum gates needed in order to prepare the DQI state, where the decoding step constitutes the most computationally expensive step for most problems. The total complexity/runtime $c_{Total}$ needed to find an optimal solution to the max-XORSAT problem at hand is then given by
\begin{equation}
    c_{Total} = c_{Opt}\cdot c_{DQI}.
\end{equation}
\subsection{Approximate benchmarking}\label{subsec:approximate_benchmarking}
\label{sec_Approx}

In practice, it is computationally infeasible to determine $p_{Opt}$ classically for larger problem instances under the assumption that the degree $l$ of the polynomial grows with $n$ respectively $m$. This is due to the prohibitive number of errors which need to be decoded. Therefore, an approximation of $p_{Opt}$ is desirable, where we do not need to decode every error of Hamming weight up to $l$. 

Our approach is to estimate the decoding failure rates $\epsilon_k$, $1\leq k\leq l$, by some sort of Monte Carlo procedure, where we just pick a fixed number (for the results of the present article $2000$) of random errors of Hamming weight $k$, $1\leq k\leq l$, decode them and take the arithmetic mean of the failure rate of the decoder on these errors $\tilde{\epsilon}_k$ as an approximation for the real failure rates  $\epsilon_k$ of the decoder. While this procedure leads to good approximations of the failure rates, it unfortunately does not suffice to compute $p_{Opt}$, as it does not provide any insights on where the wrongly decoded errors are positioned, i.e. whether they are weighted positively or negatively in the inner sum of the right-hand side of \eqref{probability_assignment}. 

To remedy this situation, further assumptions on the distribution of decoding failures in a given Hamming shell are required. To see this, we first take a closer look again at the situation with perfect decoding: Here, the probability of measuring a given assignment $x\in\mathbb{F}_2^n$ only depends on its objective value in terms of the max-XORSAT objective function 
\begin{equation}
f(x)=\sum_{j=1}^m f_j(b_j\cdot x)=\sum_{j=1}^m(-1)^{v_j+b_j\cdot x}.    
\end{equation}
We recall that in absence of decoding failures the probability density $p(x)$ of an assignment $x\in\mathbb{F}_2^n$ is given as
\begin{equation}
     p(x)=\frac{1}{2^n}\sum_{k=0}^l \frac{w_k^2}{\binom{m}{k}}\bigg(\sum_{y\in \mathcal{E}_k} (-1)^{v\cdot y}(-1)^{(B^\intercal y) \cdot x}\bigg)^2.
\end{equation}
We note that the only part which actually depends on the assignment $x$ is the innermost sum which can be rewritten as
\begin{equation}\label{identity_elementary_symmetric}    \sum_{y\in \mathcal{E}_k} (-1)^{v\cdot y}(-1)^{(B^\intercal y) \cdot x} = \sum_{y\in \mathcal{E}_k} \quad \prod_{j\in \{1,\dots ,m\}:\, y(j)=1} f_j(b_j\cdot x),
\end{equation}
cf. \cite{Jordan_2025}[Sec. 8.1.1].
The value of the expression on the right-hand side of \eqref{identity_elementary_symmetric} can be inferred from the objective value $f(x)$ of the assignment $x$ or more precisely the number of constraints $s(x)=2f(x)-m$ satisfied by the assignment $x$. We see by induction that
\begin{equation}
    \sum_{y\in \mathcal{E}_k} \quad \prod_{j\in \{1,\dots ,m\}:\, y(j)=1} f_j(b_j\cdot x) = 2\Bigg[\sum_{0\leq 2j \leq \min(k, m-s(x))} \binom{s(x)}{k-2j}\binom{m-s(x)}{2j}\Bigg]-\binom{m}{k} =: \mathcal{A}_{k,s(x)},
\end{equation}
noting that $\mathcal{A}_{k,s(x)}$ only depends on $s(x)$ and no other properties of $x$.

\begin{figure}[H]
    \centering  
    \includegraphics[width=0.7\linewidth]{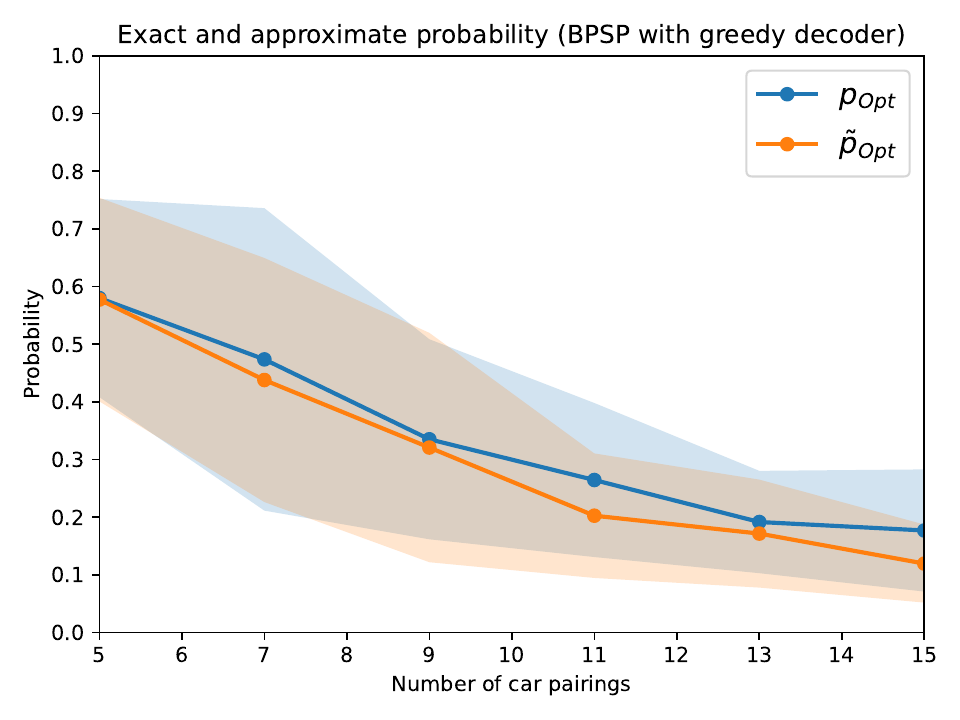}
    \caption{Testing the approximation scheme for the benchmarking procedure with the BPSP. For each number of car pairings $\{5,7,\dots, 15\}$, we randomly generate $10$ BPSP instances and compute $p_{Opt}$ and $\tilde{p}_{Opt}$. The BPSP and the greedy decoder are explained in detail in Section \ref{sec_BPSP}. The lighter areas correspond to the standard deviation of the obtained results.}
    \label{Fig_Approximation}
\end{figure}

We return to the case of imperfect decoding. We have, through approximation of the failure rates $\epsilon_k$, an approximation for the magnitude $|\mathcal{D}_k|$ of the set $\mathcal{D}_k$ of correctly decoded errors, as $|\mathcal{D}_k|=\epsilon_k \mathcal{E}_k$. However, we do not know for which $y\in\mathcal{E}_k$ we actually have $y\in\mathcal{D}_k$. In fact, this is highly dependent on the considered max-XORSAT instance. In the present article we deal with this situation by assuming that the correctly decoded errors $\mathcal{D}_k$ are distributed uniformly among their Hamming shell $\mathcal{E}_k$, in the sense that we replace the innermost sum of the right-hand side of \eqref{probability_assignment} by the arithmetic mean of all possible failure locations in the Hamming shell, i.e. by the approximation
\begin{equation}
    \frac{1}{\binom{|\mathcal{E}_k|}{|\mathcal{D}_k|}} \sum_{M\subset \mathcal{E}_k, \, |M|=|\mathcal{D}_k|} \ \sum_{y\in\mathcal M} (-1)^{v\cdot y}(-1)^{(B^\intercal y)\cdot x}.
\end{equation}
Now, this assumption is certainly not optimal and does not capture the distributions of decoding failures encountered in the actual problem instances in the present article accurately. However, some assumption is necessary at this point in order to simplify the computation, akin to the expectation value over the possible target vectors $v$ in \cite{Jordan_2025}[Thm. 10.1]. 
As seen in figure \ref{Fig_Approximation}, this estimate serves us well as a somewhat conservative estimate for the performance of DQI on some smaller BPSP instances where it is still feasible to compute $p_{Opt}$ explicitly.
Further, it is easy to work with, as one sees by induction that
\begin{equation}
    \frac{1}{\binom{|\mathcal{E}_k|}{|\mathcal{D}_k|}} \sum_{M\subset \mathcal{E}_k, \, |M|=|\mathcal{D}_k|} \sum_{y\in\mathcal M} (-1)^{v\cdot y}(-1)^{(B^\intercal y)\cdot x} = (1-\epsilon_k)\mathcal{A}_{k,s(x)}.
\end{equation}
Akin to \eqref{probability_assignment}, we find an approximate probability density $\tilde{p}_D(x)$ for a given assignment $x\in\mathbb{F}_2^n$ as
\begin{equation}
\tilde{p}_D(x):=\tilde{p}_D(s(x)):=\frac{1}{\tilde{R}}\frac{1}{2^n}\sum_{k=0}^l \frac{w_k^2}{\binom{m}{k}}\big((1-\tilde{\epsilon}_k) \mathcal{A}_{k,s(x)}\big)^2,
\end{equation}
where we recall that $\tilde{\epsilon}_k$ is our approximation of the failure rate for errors of Hamming weight $k$ and we define $\tilde{R}:=\sum_{k=0}^l w_k^2(1-\tilde{\epsilon}_k)$.
As $\tilde{p}_D(x)$ only depends on $x$ through the number of satisfied constraints $s(x)$ (as in the case with perfect decoding), the approximate probability $\tilde{p}_{Opt}$ of obtaining an optimal solution upon measuring the DQI state is then simply given as
\begin{equation}
    \tilde{p}_{Opt}= \sum_{x\in S_{Opt}}\tilde{p}_D(x) = |S_{Opt}|\ \tilde{p}_D(s_{Opt}),
\end{equation}
where $s_{Opt}$ is the number of constraints satisfied by any of the optimal solutions. The estimated expected number of times $\tilde{c}_{Opt}$ we need to run the DQI algorithm in order to find an optimal solution and the estimated total number of required quantum gates $\tilde{c}_{Total}$ are defined as 
\begin{equation}
    \tilde{c}_{Opt}=\frac{1}{\tilde{p}_{Opt}}; \qquad \tilde{c}_{Total} = \tilde{c}_{Opt}\cdot c_{DQI},
\end{equation}
in the same manner as in the section on exact benchmarking.
\section{DQI for the Binary Paint Shop Problem}\label{sec_BPSP}
\subsection{Binary Paint Shop Problem}

The Binary Paint Shop Problem (BPSP) is a combinatorial optimization problem which models a (very simple) paint shop in the automobile industry. Given a natural number $N\in\mathbb{N}$ and a set $M_N = \{m_1,\dots, m_N\}$ of $N$ distinct cars, suppose that a finite sequence $C_N=(c_1,\dots c_{2N})$ of cars in $M_N$ arrives at a car paint shop with the property that each individual car is contained exactly twice in the sequence, i.e. we have $\forall j\in\{1,\dots,2N\}:\exists! k\in\{1,\dots,2N\}: \, j\neq k \ \text{and} \ c_j=c_k$. The objective of the BPSP is to assign a sequence of colors $x=(x_1,\dots,x_{2N})\in\mathbb{F}_2^{2N}$ to the cars such that each car is painted exactly once in each of the two colors $\{0,1\}$ and such that the number of paint swaps, i.e. the $j\in\{1,\dots,2N-1\}$ with $x_j\neq x_{j+1}$, is minimized. The BPSP can hence be written as: 
\begin{align}\label{BPSP}
\min_{x \in \{0,1\}^{2N}} &\sum_{j=1}^{2N-1}1_{x_j\neq x_{j+1}} 
   \quad \text{subject to:}\\
   &\ x_j \neq x_k \ \text{if} \ c_j = c_k \ \text{and} \ j\neq k,
\end{align}
where $1_{x_j\neq x_{j+1}}:=\begin{cases}
1, \ \text{if} \ x_j \neq x_{j+1} \\ 0, \ \text{else}    
\end{cases}.$ 

The fact that both the objective function and the constraints in \eqref{BPSP} can be written as max-XORSAT constraints leads to a straightforward formulation of the BPSP in max-XORSAT with the $3N-1$ constraints:
\begin{align}\label{max_XORSATnoICC}
    x_j \oplus x_{j+1} = 0, \ \forall j\in\{1,\dots,2N-1\}, \qquad x_j \oplus x_k = 1, \ \forall j,k\in\{1,\dots,2N\}: \, c_j=c_k,
    \end{align}
where $\oplus$ denotes addition in $\mathbb{F}_2$. However, in this way we treat the terms in the objective function and the constraints from \eqref{BPSP} in the same manner, namely as max-XORSAT constraints. Therefore, solutions obtained from measuring the DQI state corresponding to the max-XORSAT problem given by \eqref{max_XORSATnoICC} might not satisfy all the required `hard` constraints and therefore do not constitute valid solutions to the original BPSP. Unfortunately, the straightforward idea of weighting the constraints in \eqref{max_XORSATnoICC} differently to avoid this problem is not compatible with the DQI algorithm. Directly weighting the constraints is not possible, as the application of Hadamard transform in the last step of the DQI state preparation would no longer lead to the desired DQI state. Repeating the same constraint multiple times is also impractical, as this would lead to substantially more decoding failures: Repeating the same constraint leads to multiple errors which map to the same syndrome (i.e. a code distance of 2); but only one of these errors can be decoded correctly from the single syndrome. Still, it is possible to prepare the DQI state, measure it, and simply sort out obtained solutions which are infeasible. However, it has proven more in practice to circumvent this problem entirely by encoding the BPSP differently in max-XORSAT - namely via the initial car color (ICC) encoding.

As each car is painted in exactly two colors, it suffices to assign the color chosen for the initial time the car arrives at the paint shop. The color chosen for the second occurrence of this car is then already uniquely determined by the initial color. Hence, it suffices to have a sequence of colors $x=(x_1=x(m_1),\dots,x_N=x(m_N))$ corresponding to the $N$ distinct cars in $M_N$. The BPSP problem then reduces to finding such an $x$ such that the objective function
\begin{equation}
    \sum_{j=1}^{2N-1}x(c_j)\oplus 1_{c_j\in\{c_1,\dots,c_{j-1}\}} \oplus x(c_{j+1})\oplus 1_{c_{j+1}\in\{c_1,\dots,c_{j}\}}
\end{equation}
is minimized. This can also be written in terms of a max-XORSAT problem - now with $m = 2N-1$ constraints:
\begin{equation}\label{max_XORSATICC}
    x(c_j)\oplus x(c_{j+1}) =  1_{c_j\in\{c_1,\dots,c_{j-1}\}} \oplus 1_{c_{j+1}\in\{c_1,\dots,c_{j}\}}, \ j\in\{1,\dots,2N-1\}.
\end{equation}
Giving rise to a constraint matrix $B\in\mathbb{F}_2^{(2N-1)\times N}$ and a vector $v\in\mathbb{F}_2^{2N-1}$ given as
\begin{align}\label{B_v_ICC}
    B_{j,k} &:= \begin{cases}
        1, \ \text{if} \ m_k =c_j \ \text{or} \ m_k = c_{j+1} \\
        0, \ \text{else}
    \end{cases}, \qquad \ \ j\in\{1,\dots,2N-1\} \notag \\ v_j &:= 1_{c_j\in\{c_1,\dots,c_{j-1}\}} \oplus 1_{c_{j+1}\in\{c_1,\dots,c_{j}\}}, \quad\qquad j\in\{1,\dots,2N-1\}.
\end{align}
In order to evaluate the performance of DQI on the BPSP it is very important to consider the quality of the LDPC code corresponding to the chosen max-XORSAT encoding of the BPSP. Our first step towards evaluating this quality is to determine the code distance $d^\bot$ of the LDPC code, i.e. the Hamming weight of the lowest-weight non-zero codeword of the code. However, before we do so, we slightly improve the quality of the code through elimination of variables respectively constraints in our max-XORSAT encoding of the BPSP. 

We consider two cases where we are able to eliminate constraints. The first case occurs when a car is repeated directly in the car sequence, i.e. if there exists $j_0\in\{1,\dots,2N-1\}$ such that $c_{j_0}=c_{j_0+1}$. Then a paint swap needs to occur between $c_{j_0}$ and $c_{j_0+1}$. Hence, we can eliminate the constraint $x_{j_0}\oplus x_{j_0+1}=0$ in \eqref{max_XORSATnoICC}, respectively, the constraint $x(c_{j_0})\oplus x(c_{j_0+1})=v_{j_0}$ in \eqref{max_XORSATICC}.

The second case occurs when a car is repeated with exactly one other car in between, i.e. if there exists  $j_0\in\{1,\dots,2N-2\}$ such that $c_{j_0}=c_{j_0+2}$. Then exactly one paint swap needs to occur $c_{j_0}$ and $c_{j_0+2}$. Where this paint swap occurs precisely (whether between $c_{j_0}$ and $c_{j_0+1}$ or between $c_{j_0+1}$ and $c_{j_0+2}$) can be determined by the position of the other paint swaps. Therefore, we are able to eliminate the constraints $x_{j_0}\oplus x_{j_0+1}=0$ and $x_{j_0+1}\oplus x_{j_0+2}=0$ in \eqref{max_XORSATnoICC}, respectively, the constraints $x(c_{j_0})\oplus x(c_{j_0+1})=v_j$ and $x(c_{j_0}+1)\oplus x(c_{j_0+2})=v_{j_0+1}$ in \eqref{max_XORSATICC}. Through this elimination of constraints it possible that one of the variables $x_j$, $j\in\{1,\dots, 2N\}$ resp. $j\in\{1,\dots, N\}$, is no longer contained in any of the constraints. In this case we also eliminate the variable in question.

To illustrate this procedure we consider the example BPSP given by
\begin{equation}\label{BPSP_example}
    N=5, \qquad M_N=\{1,2,3,4,5\}, \qquad C_N = (1,2,1,3,4,5,2,5,3,4),
\end{equation}
and determine its ICC encoding. Applying the definition in \eqref{B_v_ICC} we have
\begin{equation}\label{max_XORSAT_example_orig}
     B = \left(\begin{array}{ccccc}
       1  & 1 & 0 & 0 & 0 \\
       1  & 1 & 0 & 0 & 0 \\
       1 & 0 & 1 & 0 & 0 \\
       0 & 0 & 1 & 1 & 0  \\
       0 & 0 & 0 & 1 & 1 \\
       0 & 1 & 0 & 0 & 1 \\
       0 & 1 & 0 & 0 & 1 \\
       0 & 0 & 1 & 0 & 1 \\
       0 & 0 & 1 & 1 & 0 
    \end{array}\right), \qquad v= \left(\begin{array}{c}
       0  \\
       1  \\
       1  \\
       0  \\
       0  \\
       1  \\
       0  \\
       0 \\
       0
    \end{array}\right),
\end{equation}
before eliminating any constraints. As $c_1=c_3$ and $c_6=c_8$, we eliminate the first, the second, the sixth, and the seventh constraint, being left with
\begin{equation}\label{max_XORSAT_example_reduced}
     B = \left(\begin{array}{ccccc}
       1 & 0 & 1 & 0 & 0 \\
       0 & 0 & 1 & 1 & 0  \\
       0 & 0 & 0 & 1 & 1 \\
       0 & 0 & 1 & 0 & 1 \\
       0 & 0 & 1 & 1 & 0 
    \end{array}\right), \qquad v= \left(\begin{array}{c}
       1  \\
       0  \\
       0  \\
       0 \\
       0
    \end{array}\right).
\end{equation}
As $x_2$ is not part of any of the remaining constraints, we eliminate it as a variable. We then consider the reduced problem with variables $x=(x_1,x_3,x_4,x_5)\in\mathbb{F}_2^4$ and the constraints
\begin{equation}\label{constraints_reduction_final}
    x_1 \oplus x_3 = 1, \quad x_3 \oplus x_4 = 0, \quad x_4 \oplus x_5 = 0, \quad x_5 \oplus x_3 = 0, \quad  \text{and} \quad x_3\oplus x_4 =0.
\end{equation}
One of the optimal solutions with respect to the constraints \eqref{constraints_reduction_final} is $x_1=0$, $x_3=1$, $x_4=1$ and $x_5=0$. The choice of $x_2$ does not matter for the number of satisfied constraints, respectively, number of paint swaps. Choosing, e.g., $x_2=0$, we recover an optimal solution to the original non-ICC encoded BPSP as $(x_1,\dots,x_{10})=(0,0,1,1,1,1,1,0,0,0)$ with a total of $2$ paint swaps.

We now turn to the distance of the LDPC code generated by the max-XORSAT encodings of the BPSP. We recall that the code $C^\bot=\{d\in\mathbb{F}_2^m: \, B^\intercal d=0\}$ corresponding to a max-XORSAT problem is the LDPC code with parity-check matrix $B^\intercal$. The minimum distance $d^\bot$ of this code is defined as the minimum Hamming weight of any nonzero codeword $d\in C^\bot$. Alternatively, $d^\bot$ can be computed as the minimum number of linearly dependent columns of the parity-check matrix $B^\intercal$ (respectively rows of $B$). Hence, we see both the original max-XORSAT encoding \eqref{max_XORSAT_example_orig} and the reduced encoding \eqref{max_XORSAT_example_reduced} of our BPSP example \eqref{BPSP_example} have a code distance of $2$, i.e. the elimination of constraints does not necessarily lead to an increased code distance in the ICC encoding. This stands in contrast to the non-ICC encoding where one finds
\begin{equation}
    d^\bot=\min_{\substack{k\in\{1,\dots,2N\}:\\ \, \exists j\in\{1,\dots, 2N-k\}: \, c_j=c_{j+k}}}(k+1)
\end{equation}
without variable elimination. After eliminating constraints, all occurrences of the same car have a minimum distance of $3$ with respect to the constraints. Therefore, we have $d^\bot\geq 4$ in this case.

\begin{figure}[H]
    \centering  
    \includegraphics[width=0.45\linewidth]{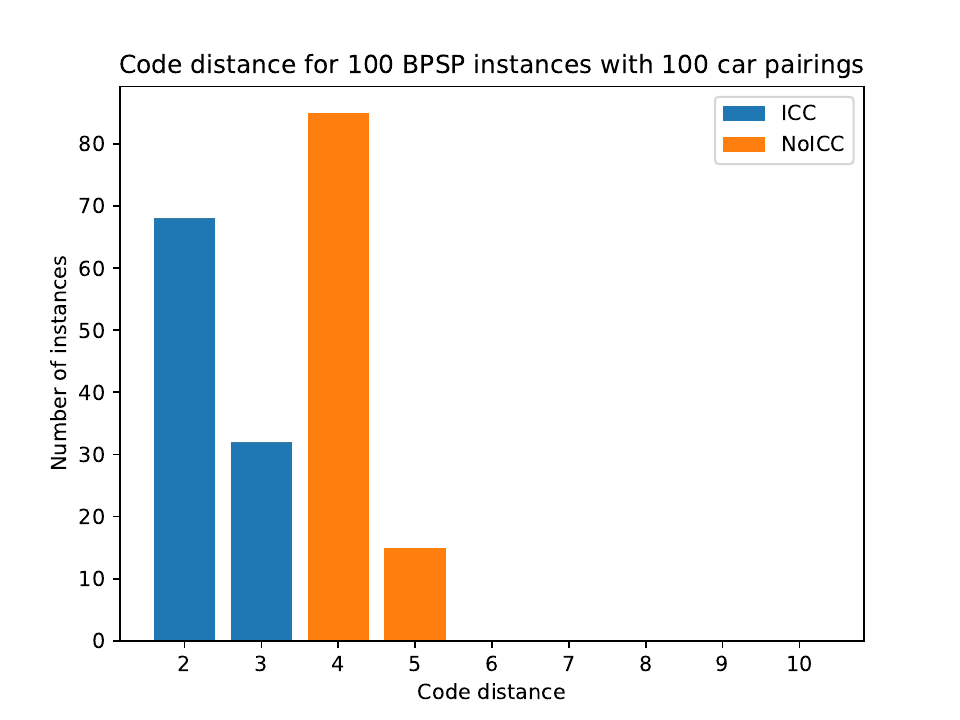}
    \includegraphics[width=0.45\linewidth]{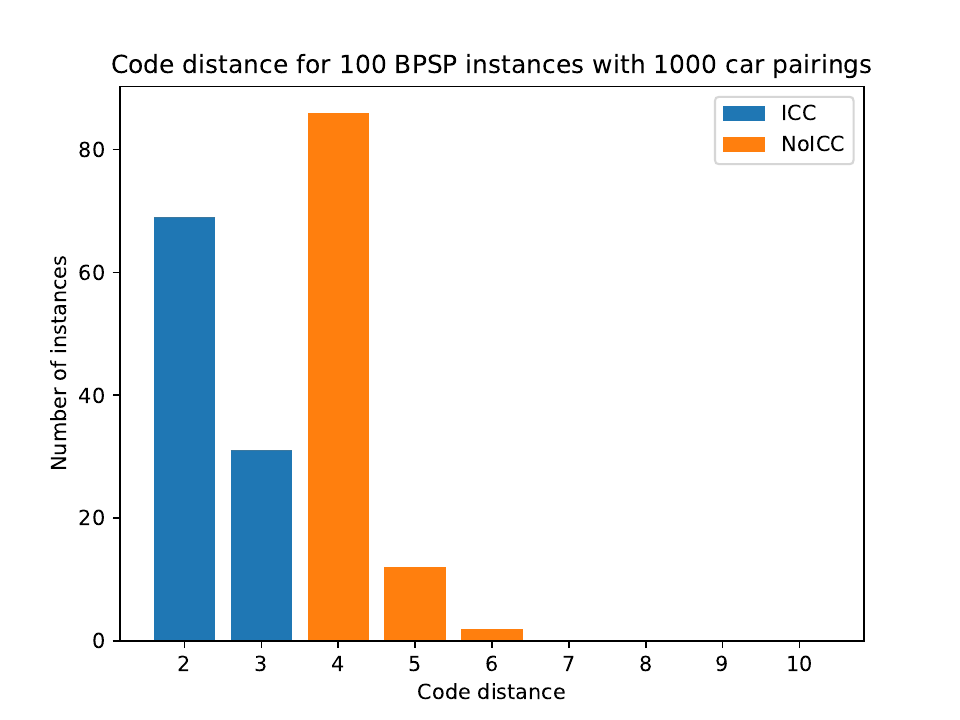}
    \caption{Code distance for a number of randomly generated BPSP instances. For both $100$ and $1000$ car pairings, we randomly generate $100$ instances each and compute the corresponding code distances $d^\bot$. We note that the code distance $d^\bot$ does not increase meaningfully with the higher number of car pairings.} 
    \label{Fig_code_dist}
\end{figure}

As we see in Fig.~\ref{Fig_code_dist}, the distance of the LDPC code is essentially constant in the sense that it essentially does not grow with the size of the BPSP given by $n$, respectively, $m$. Therefore, the fact that \cite{Jordan_2025}[Thm. 10.1] assumes a polynomial degree $l$ linear in $m$ suggests that we need to decode somewhat reliably well beyond the code distance $d^\bot$ in order to obtain promising results. This is also in agreement with the findings in \cite{Parekh_2025} on MaxCut (which is a special case of max-2-XORSAT considered here) that good results under the assumption of perfect decoding are only obtainable for a specific class of high-girth graphs where efficient exact classical solvers exist. 

Further, we see that choosing the ICC encoding is indeed a tradeoff, as this leads to a lower distance of the LDPC code and hence to a more failure-prone decoding step. Still, from our computer experiments the advantages of the ICC encoding seem to outweigh the lower code distance in practice, which is why we restrict ourselves to the ICC encoding for the remaining part of the present article.

\subsection{Decoder for the BPSP}\label{subsec:bpsp_decoder}
We now turn towards our choice of decoder for the decoding step in the DQI algorithm. We first note that our max-XORSAT encodings for the BPSP from the previous section are actually max-2-XORSAT problems, i.e. each constraint contains at most (in the BPSP case even exactly) $2$ variables. This allows reformulation of the decoding problem in terms of a graph-theoretic problem. Identifying the columns of the matrix $B$ with a set of vertices $V=\{v_1,\dots, v_n\}$ and the rows of $B$ with a set of edges $E=\{e_1,\dots e_m\}$, we obtain a connected graph $G=(V,E)$ where each edge $e_j\in E$ connects precisely these two vertices $v_k,v_l\in V$ which are contained in the constraint corresponding to $e_j$, i.e.~for which we have $B_{jk}=1=B_{jl}$. Therefore, the problem of finding a given error $y\in\mathbb{F}_2^m$ from the corresponding syndrome $B^\intercal y\in\mathbb{F}_2^n$ can be reformulated as finding a subset $U\subseteq E$ from the even-magnitude syndrome subset $T:=\{v_k\in V: \, B^\intercal y_k=1\}$. The decoded error $\tilde{y}\in\mathbb{F}_2^m$ can then be constructed from $U$ by setting $\tilde{y}_k=1$ if and only if $e_k\in U$. As the inclusion of a given edge $e_j\in E$ in $U$ corresponds to flipping two components (associated with the vertices connected by the edge) of the corresponding syndrome $B^\intercal \tilde{y}$, we have $(B^\intercal \tilde{y})_k= \deg_{(V,U)}(v_k)$, where $\deg_{(V,U)}(v_k)\in \mathbb{F}_2$ denotes the (even or odd) degree of the $k$th vertex $v_k$ in the subgraph $(V,U)$ of $G$. Therefore, the subset of edges $U$ corresponds to an error $\tilde{y}$ that correctly maps to the syndrome corresponding to $T$ if and only if $U$ is a $T$-join of $G$, defined in
\begin{definition}
    Let $G=(V,E)$ be a graph and $T\subseteq V$ a subset of vertices. A $T$-join $U\subseteq E$ of $G$ is a subset of edges such that $T$ is precisely the set of vertices of odd degree in the subgraph $(V,U)$ of $G$.
\end{definition}
In the decoding step, one is interested in finding a minimum weighted $T$-join, that is a $T$-join $U$ such that $|U|$ is minimal. This corresponds to the Hamming weight of the decoded error being minimal, i.e. the decoder being a minimum length decoder. We note that a minimum weighted $T$-join is not unique. Indeed, as observed before, we consider max-XORSAT instances where the corresponding LDPC code has a distance of $2$, i.e. there are multiple edges connecting a single pair of vertices in $G$. This necessitates the existence of multiple minimum weighted $T$-joins. Finding such a minimum weighted $T$-join is usually done in two steps.

First, we construct a new complete weighted graph $G'=(T,E',w)$ with the vertices given by the syndrome subset $T$, the set of edges $E'$ being complete, and the weights defined as $w(v_1,v_2):=d_G(v_1,v_2)$, $v_1, v_2\in T$, where $d_G(v_1,v_2)$ denotes the length of the shortest path from $v_1$ to $v_2$ in the original Graph $G$. We recall that in the case of the BPSP (with variable elimination) the graph $G$ is connected, in particular we have $d_G(v_1,v_2)<\infty$ for all $v_1,v_2\in T\subseteq V$. In order to construct the graph $G'$, we need to find the shortest paths between each pair of vertices in $T$. A naive approach to do so would be to run Dijkstra's algorithm on each of these pairs. In practice, there are more efficient algorithms to do so, as we discuss along the quantum circuit implementation of our decoder.

The second step is to construct a minimum-weight perfect matching of the new graph $G'$, that is a set $M\subset E'$ of pair-wise non-adjacent edges such that each vertex $v\in T$ is incident to (exactly) one of the edges in $M$ and such that the total weight $W:=\sum_{e\in M}w_e$ is minimized. Such a matching can be found, as the number of vertices in $T$ is even by construction. The most well-known algorithm to construct a minimum-weight perfect matching on weighted graphs is the appropriate weighted extension \cite{Edmonds_1965_weighted} of Edmond's blossom algorithm \cite{Edmonds_1965} which runs in time $O(|T|^4)$. With such a matching at hand, we are now able construct our minimum $T$-join $U$. Each edge $e\in M\subset E'$ corresponds to a shortest path in the original graph $G$ (if there are multiple shortest paths, we need to make a choice here in order to obtain a $1-1$-correspondence), and each of these paths consists of a subset of edges in $E$. One can show that all the paths corresponding to the matching $M$ are necessarily disjoint, given that the matching has minimum weight. The minimum weighted $T$-join $U$ is then given as the set of all edges contained in one of the paths associated to the minimum-weight perfect matching $M$. We construct the decoded error from the set $U=\{e_k\in E: \, \tilde{y}_k=1\}$.

We now turn towards our concrete implementation of such a max-2-XORSAT decoder as a quantum circuit. We deviate from the just described general procedure in two ways. Firstly, we want to separate the two steps described above into a classical precomputation step, where we construct a new graph $G'$ via the computation of the shortest paths in $G$, and an actual quantum circuit, where the information on the syndrome set $T$ enters. To facilitate this, we start by first classically constructing a complete weighted graph $G'=(V,E',w)$ which does not depend on the given syndrome. Secondly, instead of constructing a minimum weight $T$-join, we construct a $T$-join with potentially suboptimal weight by replacing the blossom algorithm by a greedy algorithm which finds a (potentially suboptimal) perfect matching. This comes with two advantages. On the one hand, the greedy algorithm is simply easier to implement as a quantum circuit than the blossom algorithm. On the other hand, the greedy algorithm runs significantly faster than the blossom algorithm. This allows us to prepare the DQI state more often in the same amount of time when using the faster decoder. This effect can outweigh the worse decoding quality of the suboptimal greedy decoder, as shown by our results. There are algorithms which construct a minimum-weight perfect matching on a weighted graph faster than the blossom algorithm. To our knowledge, the algorithm \cite{GabowTarjan1991} is the fastest algorithm available in our special case, making use of the fact that all weights are integer-valued. However, implementing this algorithm as a quantum circuit should prove difficult and is beyond the scope of the present paper.

\begin{figure}[H]
    \centering  
    \includegraphics[width=0.45\linewidth]{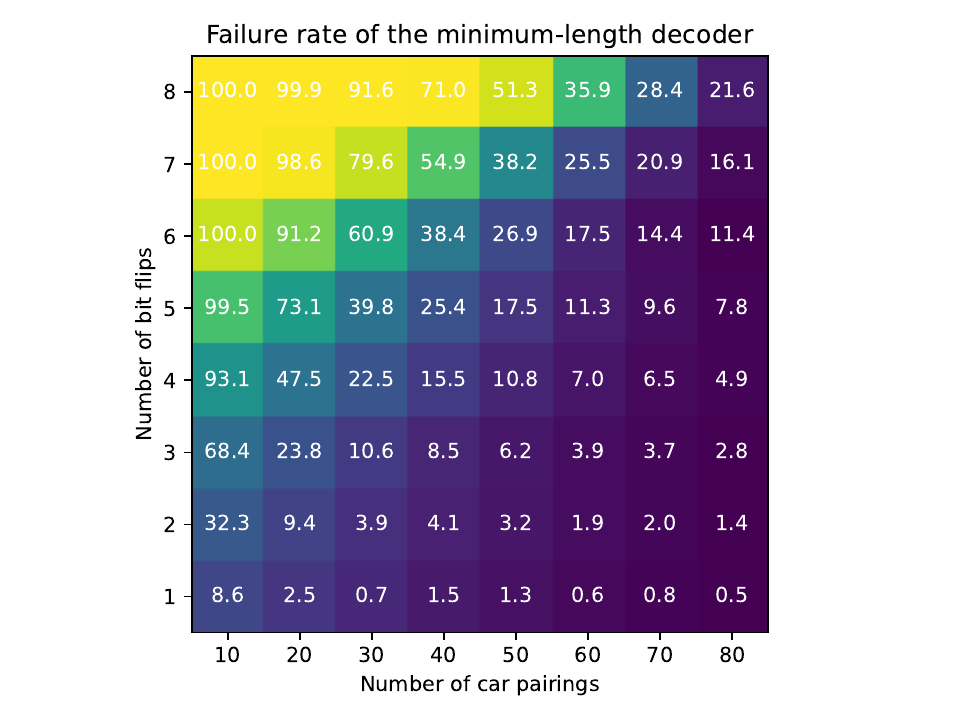}
    \includegraphics[width=0.45\linewidth]{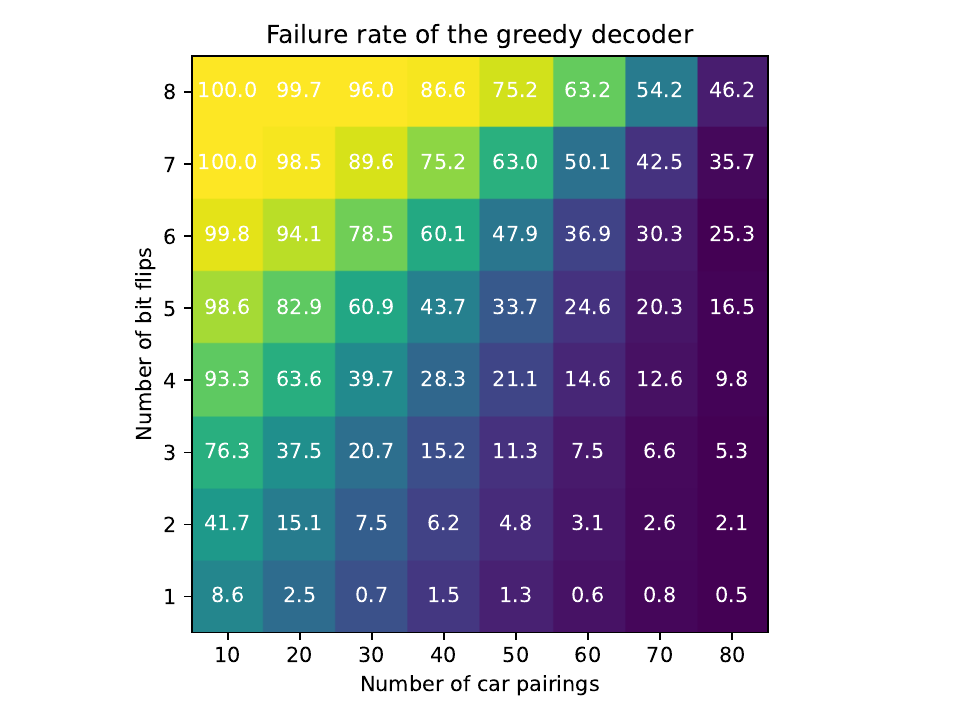}
    \caption{Comparison of failure rates for a minimum-length decoder and our implementation of the greedy decoder. For each number of car pairings, we generate $10$ random BPSP instances and decode $2000$ randomly generated errors per Hamming weight and instance.}
    \label{Fig_decoder_flips}
\end{figure}

We begin by describing the classical precomputation step. As described before, a given max-2-XORSAT problem gives rise to a graph $G=(V,E)$ through its constraint matrix $B$. Starting from $G$, we first compute the shortest paths between all pairs of vertices $v_1,v_2\in V$. As $G$ is a connected and still unweighted graph, the most efficient way to do this is given by Seidel's algorithm \cite{Seidel_1995}. It yields the distances between each pair of vertices in time $O(n^\nu \log n)$, where $\nu<2.376$ denotes the exponent of the complexity of multiplying two $n\times n$ matrices of small integers (and we recall $|V|=n$). Given the distances, the shortest paths can also be constructed in time $O(n^\nu \log n)$. The set of shortest paths $P:=\{p(v_1,v_2): \, \text{shortest path between } v_1 \ \text{and} \ v_2; \ \text{for}\ v_1,v_2\in V\} $ is then sorted by path length (and subsequently by start vertex and end vertex of the path). As $|P|=\binom{n}{2}=O(n^2)$ this can be done in time $O(n^2\log n)$. Each path $p\in P$ corresponds to a finite sequence of edges in the original graph $G$. For a given edge $e\in E$, we write $e\in p$ if the edge is part of the path $p$.
With the sorted list of paths at hand, we can now describe our greedy decoder:

\begin{algorithm}
    \caption{Greedy decoder}
    \begin{algorithmic}[1]
    \State \textbf{Input:} 
    An ordered list of paths $P$ in the graph $G$ corresponding to a given max-2-XORSAT problem with constraint matrix $B\in \mathbb{F}_2^{m\times n}$, an error $y\in \mathbb{F}_2^m$ and a syndrome $T=B^\intercal y\in\mathbb{F}_2^n$.
    \State \textbf{Output:} A (not necessarily correctly) decoded error $y'\in\mathbb{F}_2^m$.
    \State $y'\leftarrow y$
    \For{$p=p(v_1,v_2) \in P$}
         \If{$v_1,v_2\in T$}
            \For{$e\in p$}
               \State Flip the component of $y'$ corresponding to the edge $e$
             \EndFor   
             \State Remove $v_1$ and $v_2$ from $T$           
        \EndIf
    \EndFor
\State Return $y'$
        \end{algorithmic}
        \label{greedy_decoder}
\end{algorithm}
As the algorithm does not produce errors of minimum length, it will fail more often while decoding errors of lower Hamming weight when compared to a hypothetical minimum-length decoder using the blossom algorithm. We also note that the paths $p\in P$ for which $v_1,v_2\in T$ in algorithm \ref{greedy_decoder} do not need to be disjoint, i.e. a component of $y'$ can be flipped multiple times throughout algorithm \ref{greedy_decoder}. Comparing the performance of the hypothetical minimum-length decoder to the greedy decoder in figure~\ref{Fig_decoder_flips}, we see that the greedy decoder indeed performs worse on errors of lower Hamming weight.

In order to keep track of the iteration through the list of ordered paths $P$, we introduce an auxiliary path register of size $N \coloneqq \binom{n}{2}$.
We label its qubits $p_1, \dots, p_N$.
Further, the qubits of the syndrome register containing $\ket{B^T y}$ will be labelled $v_1,\dots,v_n$, highlighting the correspondence to the vertices of the graph $G$. 
In the same spirit, we will refer to the qubits in the error register containing $y$ by $e_1,\dots,e_m$.

The conditions in algorithm \ref{greedy_decoder} can be realized in a quantum circuit by CNOT gates respectively Toffoli gates.
We denote a CNOT gate controlling on qubit $a$ and acting on $b$ by $CX(a,b)$. 
Similarly, a Toffoli gate controlling on qubits $a$ and $b$ and acting on $c$ will be written as $CCX(a,b,c)$.

\begin{algorithm}
    \caption{Quantum circuit generation}
    \begin{algorithmic}[1]
    \State \textbf{Input:} A quantum state $\ket{B^\intercal y}\ket{0}^N \ket{y}$, a connected graph $G=(V,E)$ an ordered list of paths $P = [p_1,\dots,p_N]$ between all pairs of vertices $v,w \in V$
    \State \textbf{Output:} A quantum state $\ket{B^\intercal y}\ket{g}\ket{y'}$ where $\ket{g}$ is a $N$-qubit garbage state and $\ket{y'}$ is the (not necessarily correctly) decoded error
    \For{$p=p(v_1,v_2) \in P$}
    \State Apply $U_i \coloneqq CX(p,v_1)CX(p,v_2)\big(\prod_{e \in p}CX(p,e)\big) CCX(v_1,v_2,p)$
    \EndFor
    \For{$p=p(v_2,v_1) \in P$}
    \State Apply $CX(p,v_1)CX(p,v_2)$ \Comment{Restore syndrome register}
    \EndFor
\end{algorithmic}
        \label{quantum_circuit_generation}
\end{algorithm}

To illustrate the construction of the quantum circuit we come back to our example BPSP \eqref{BPSP_example}. After elimination of variables, we are left with $n=4$, $m=5$ and the constraint matrix 
\begin{equation}
     B = \left(\begin{array}{cccc}
       1 &  1 & 0 & 0 \\
       0 &  1 & 1 & 0  \\
       0 &  0 & 1 & 1 \\
       0 &  1 & 0 & 1 \\
       0 &  1 & 1 & 0 
    \end{array}\right).
\end{equation}
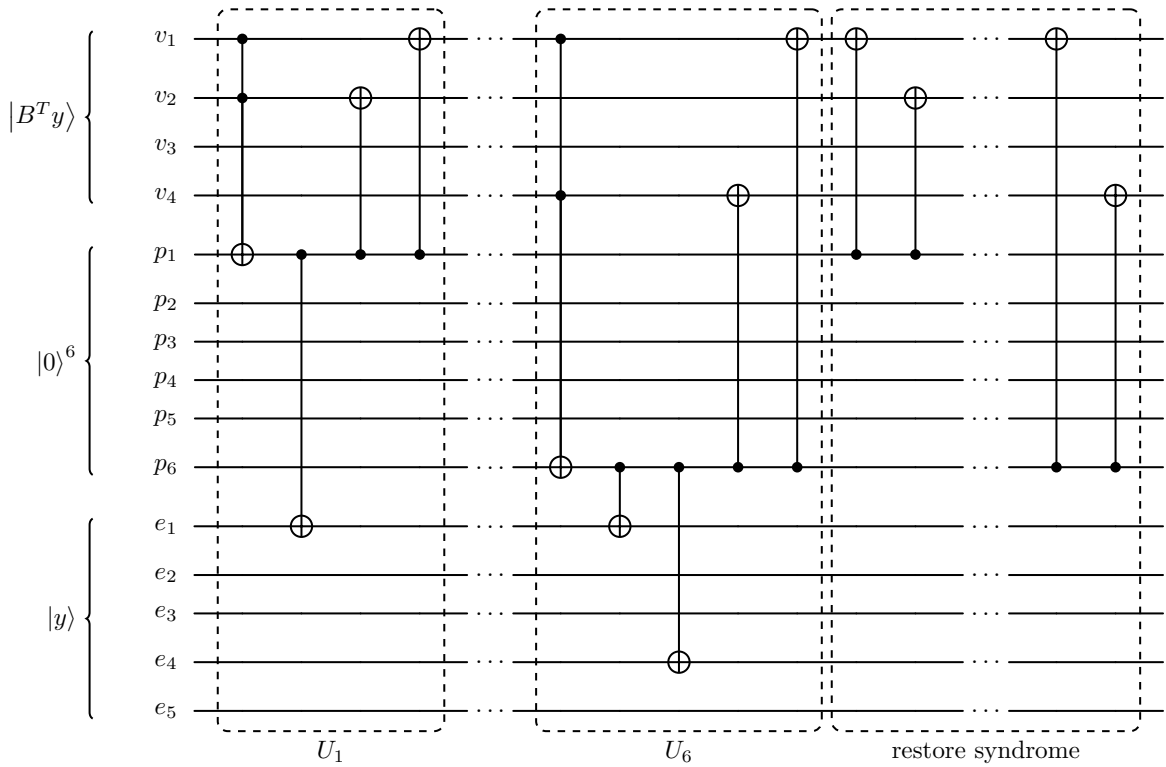
\begin{figure}
\label{fig:decoding_circuit}
\begin{adjustbox}{width=1\textwidth, nofloat = figure}
\begin{quantikz}
\lstick[4]{$\ket{B^T y}$}&\wireoverride{2} \midstick{$v_1$} & \ctrl{4}\gategroup[15,steps=4,style={dashed,rounded
corners, inner
xsep=2pt},background,label style={label
position=below,anchor=north,yshift=-0.2cm}]{$U_1$} &&&\targ{}& \ \ldots \ &\ctrl{9}\gategroup[15,steps=5,style={dashed,rounded
corners, inner
xsep=2pt},background,label style={label
position=below,anchor=north,yshift=-0.2cm}]{$U_6$}&&&&\targ{}&\targ{}\gategroup[15,steps=5,style={dashed,rounded
corners, inner
xsep=2pt},background,label style={label
position=below,anchor=north,yshift=-0.2cm}]{restore syndrome}&& \ \ldots \ &\targ{}&& \\
& \wireoverride{2} \midstick{$v_2$}&\ctrl{3}&&\targ{}&& \ \ldots \ &&&&&&&\targ{}& \ \ldots \ &&&\\
& \wireoverride{2} \midstick{$v_3$}&&&&& \ \ldots \ &&&&&&&& \ \ldots \ &&&\\
& \wireoverride{2} \midstick{$v_4$}&&&&& \ \ldots \ &\ctrl{6}&&&\targ{}&&&& \ \ldots \ &&\targ{}&\\
\lstick[6]{$\ket{0}^6$}&\wireoverride{2} \midstick{$p_1$} &\targ{} & \ctrl{6}&\ctrl{-3}&\ctrl{-4}& \ \ldots \ &&&&&&\ctrl{-4}&\ctrl{-3}& \ \ldots \ &&& \\
& \wireoverride{2} \midstick{$p_2$}&&&&& \ \ldots \ &&&&&&&& \ \ldots \ &&&\\
& \wireoverride{2} \midstick{$p_3$}&&&&& \ \ldots \ &&&&&&&& \ \ldots \ &&&\\
& \wireoverride{2} \midstick{$p_4$}&&&&& \ \ldots \ &&&&&&&& \ \ldots \ &&&\\
& \wireoverride{2} \midstick{$p_5$}&&&&& \ \ldots \ &&&&&&&& \ \ldots \ &&&\\
& \wireoverride{2} \midstick{$p_6$}&&&&& \ \ldots \ &\targ{}&\ctrl{1}&\ctrl{4}&\ctrl{-6}&\ctrl{-9}&&& \ \ldots \ &\ctrl{-9}&\ctrl{-6}&\\
\lstick[5]{$\ket{y}$}&\wireoverride{2} \midstick{$e_1$}&&\targ{}&&& \ \ldots \ &&\targ{}&&&&&& \ \ldots \ &&&\\
& \wireoverride{2} \midstick{$e_2$}&&&&& \ \ldots \ &&&&&&&& \ \ldots \ &&&\\
& \wireoverride{2} \midstick{$e_3$}&&&&& \ \ldots \ &&&&&&&& \ \ldots \ &&&\\
& \wireoverride{2} \midstick{$e_4$}&&&&& \ \ldots \ &&&\targ{}&&&&& \ \ldots \ &&&\\
& \wireoverride{2} \midstick{$e_5$}&&&&& \ \ldots \ &&&&&&&& \ \ldots \ &&&
\end{quantikz}
\end{adjustbox}
\caption{Quantum circuit for greedy decoding of the BPSP instance \eqref{BPSP_example}. The decoding part of the circuit comprises one block for each path in $P$. Subsequently, the information stored in the auxiliary register is used to restore the syndrome register.}
\end{figure}
Translating this into a graph, we obtain $G=(V,E)$ with $V=\{v_1,v_2,v_3,v_4\}$ and 
\begin{equation}
   E=\{e_1=(v_1,v_2),e_2=(v_2,v_3),e_3=(v_3,v_4), e_4=(v_2,v_4), e_5=(v_2,v_3)\}. 
\end{equation}
We note that the edge $e_5$ is incident to the same vertices as $e_2$ and is therefore redundant. Hence, we remove it when constructing the graph $G'$. In order for $G'$ to be complete we need to add two additional edges $p_5$ corresponding to the path $(e_1,e_2)$ and $p_6$ corresponding to the path $(e_1,e_4)$, both of weight $2$. We therefore obtain the ordered path list
\begin{equation}
    P=[p_1=(e_1), p_2 =(e_2), p_3=(e_4), p_4=(e_3), p_5=(e_1,e_2), p_6=(e_1,e_4)].
\end{equation}
The quantum circuit for decoding errors in this problem instance is depicted in figure~\ref{fig:decoding_circuit}.

\begin{figure}[H]
    \centering  
    \includegraphics[width=0.9\linewidth]{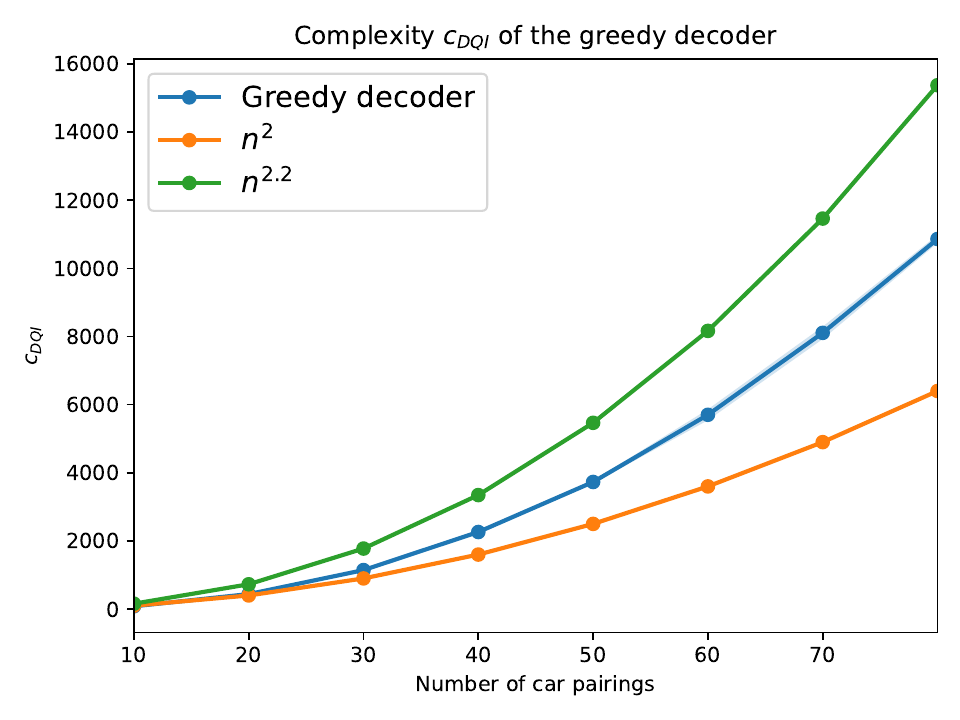}
    \caption{Leading order contribution to the number of quantum gates needed for the quantum circuit of the greedy decoder described in algorithm~\ref{quantum_circuit_generation}. We randomly generate $10$ BPSP instances for each number of car pairings. The two other curves are added to make it easier to visualize the growth rate of this leading order contribution. The light blue area around the blue graph (barely visible) corresponds to the standard deviation of the distribution of the obtained results.}
    \label{Fig_greedy_decoder}
\end{figure}

For the purpose of benchmarking our DQI implementation, we are also interested in the performance of the greedy decoder. The leading order contribution to the number of quantum gates in the decoder described by algorithm \ref{greedy_decoder} stems from the CNOT gates which are controlled on the path register and flip the error register. As each path $p\in P$ contributes as many of these CNOT gates as its length, the number of these gates grows faster than the number $\binom{n}{2}=|P|$ of paths and corresponding Toffoli and CNOT gates. This leading order contribution to the number of gates also corresponds to the sum of all edge weights in the completed graph $G'$. The concrete scaling heavily depends on the chosen instances of max-2-XORSAT problems. For our implementation of the BPSP, the scaling is heuristically analyzed in figure~\ref{Fig_greedy_decoder}.

\section{Results}\label{sec:results}

For our benchmarking results on the BPSP, we randomly generate $10$ BPSP instances for each number of car pairings $\{10,20,\dots,70,80\}$. Due to the higher number of car pairings involved we are only able to perform the approximate benchmarking described in Section~\ref{sec_Approx}. We compute the approximate probability $\tilde{p}_{Opt}$ of measuring an optimal solution for the DQI state for each individual instance. To do so, we compute the set of all optimal solutions for each instance with the help of Gurobi. We do this for both our implementation of the greedy decoder, described in Section~\ref{sec_BPSP}, and a hypothetical minimum-length decoder utilizing the blossom algorithm. We stress the fact that implementing a quantum circuit for the blossom algorithm is beyond the scope of the present article and therefore all results concerning the blossom decoder are hypothetical, based on the fact that we are still able to evaluate the results classically. 

\begin{figure}[H]
    \centering  
    \includegraphics[width=0.9\linewidth]{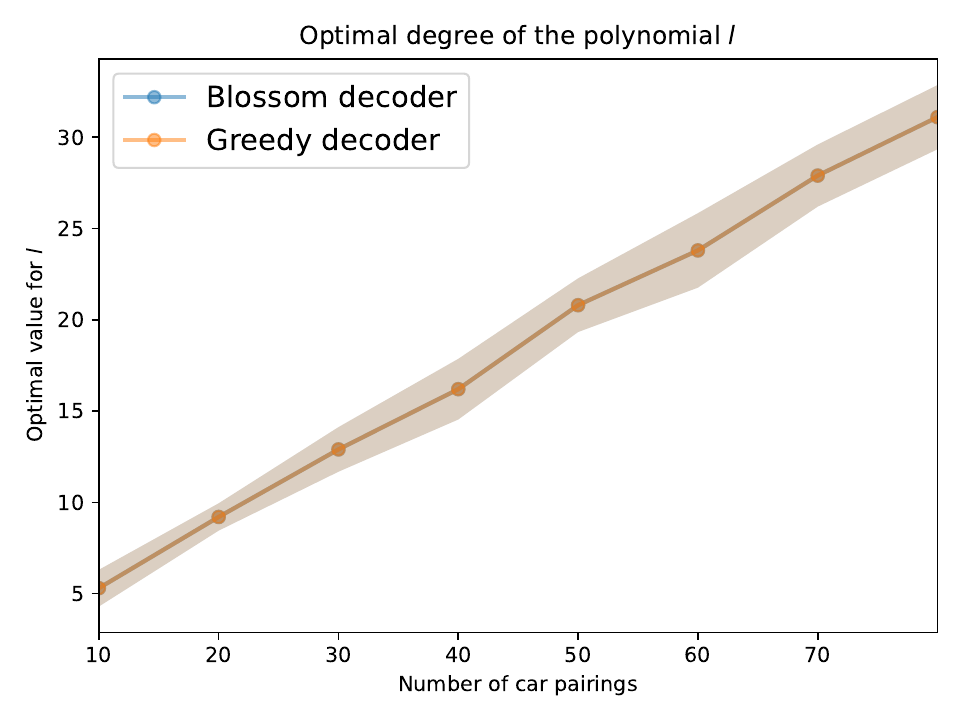}
    \caption{Optimal choice of degree for the DQI polynomial.
    For both the exact blossom decoder and our approximate greedy decoder, we approximate the probability $\tilde{p}_{Opt}$ of measuring an optimal solution to one of the given BPSP instances. By doing this for all reasonable choices of polynomial degree $0 \leq l \leq n$, we find degrees for which $\tilde{p}_{Opt}$ is optimal. Curiously, the optimal degrees coincide for both decoders on all instances even though the failure rates are quite different, cf.~figure~\ref{Fig_decoder_flips}. For the tested instances the optimal degree seems to grow roughly linearly in $n$ with slope $~\frac{2}{5}$. The lighter area corresponds to the standard deviation of the distribution of the obtained results.}
    \label{Fig_opt_deg}
\end{figure}

In our first result, depicted in figure~\ref{Fig_opt_deg}, we compute $\tilde{p}_{Opt}$ for all sensible choices of polynomial degrees $1\leq l \leq n$ and use this to determine the value of $l$ for which $\tilde{p}_{Opt}$ is optimal. We note that for this $l$ the total runtime $\tilde{c}_{Total}$ is also optimal, as the individual runtime of the decoder does not depend on $l$.  

Curiously, we obtain the same optimal polynomial degrees for both decoders on all instances, even though one might be inclined to assume that higher $l$ could be advantageous for the greedy decoder, where correctly decoded errors have more than minimum length. For both decoders, the optimal degree $l$ substantially exceeds the constant code distance $d^\bot$ (cf.~figure~\ref{Fig_code_dist}), at least on the somewhat limited instances with up to $80$ car pairings we were able to analyze. In fact, the optimal value for $l$ seems to almost grow linearly in the number of car pairings. This is in agreement with the limiting results proved in \cite{Jordan_2025}[Thm.~4.1, Thm.~10.1], where a linear growth of $l$ in the problem size is needed to sufficiently boost the expected number of satisfied constraints for the DQI state. 

\begin{figure}[H]
    \centering  
    \includegraphics[width=0.9\linewidth]{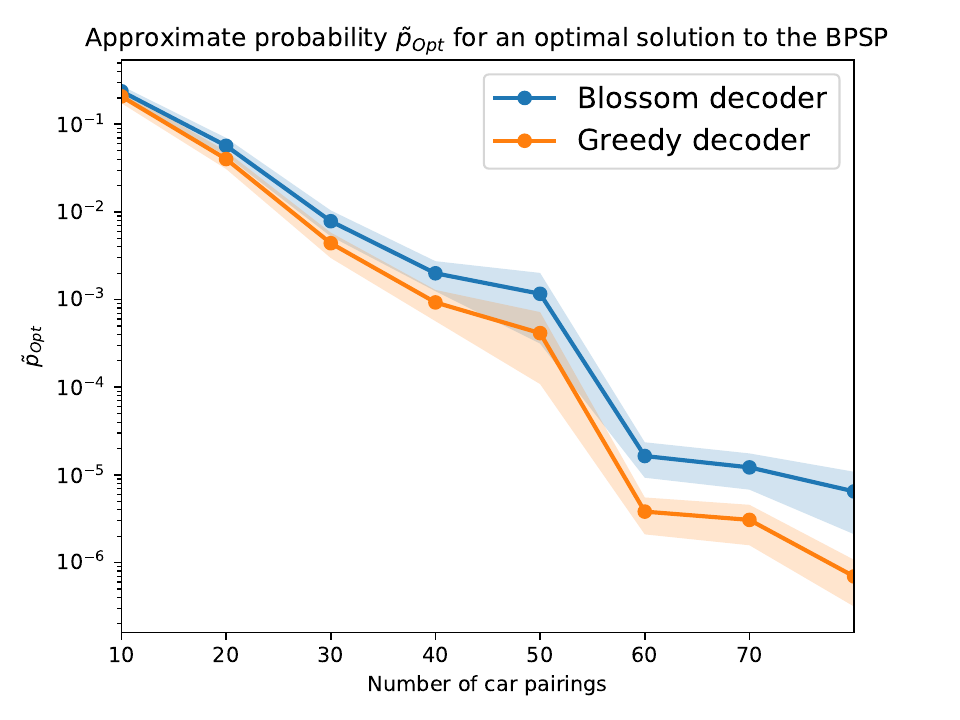}
    \caption{We compare the performance of our implementation of the greedy decoder and the hypothetical minimum-length blossom decoder on the instances described in the beginning of the section. We classically compute the approximate probability $\tilde{p}_{Opt}$ for both decoders and plot it on a logarithmic scale. As expected, the minimum length decoder beats the greedy decoder on all instances, as we do not take into account the different runtimes of the decoders. The light areas correspond to one third of the standard deviation.}
    \label{Fig_decoder_comp}
\end{figure}

With the analysis of the optimal polynomial degrees at hand, we can make an informed choice of $l$ for our benchmarking results. In accordance with figure~\ref{Fig_opt_deg}, we choose $l=\lfloor \frac{2}{5}n \rfloor $ as the largest integer smaller or equal than $\frac{2}{5}n$. Next, we compare the performance of the greedy decoder to the performance of the minimum-length decoder in figure~\ref{Fig_decoder_comp} by comparing the probability of measuring an optimal solution for both DQI states.

As expected, the blossom decoder beats the greedy decoder on all instances. Both decoders have a significantly higher probability of measuring an optimal solution than simple random sampling. Computing $\tilde{p}_{Opt}$ does not yet take into account the different runtimes of the decoders. The faster runtime of the greedy decoder (a factor at least of order $n^{1.8}$) seems be more than enough to compensate for the lower probability to measure an optimal solution as seen in figure~\ref{Fig_decoder_comp}. To see this, we next compute $\tilde{c}_{Total}$ for both of the decoders.

\begin{figure}[H]
    \centering  
    \includegraphics[width=0.9\linewidth]{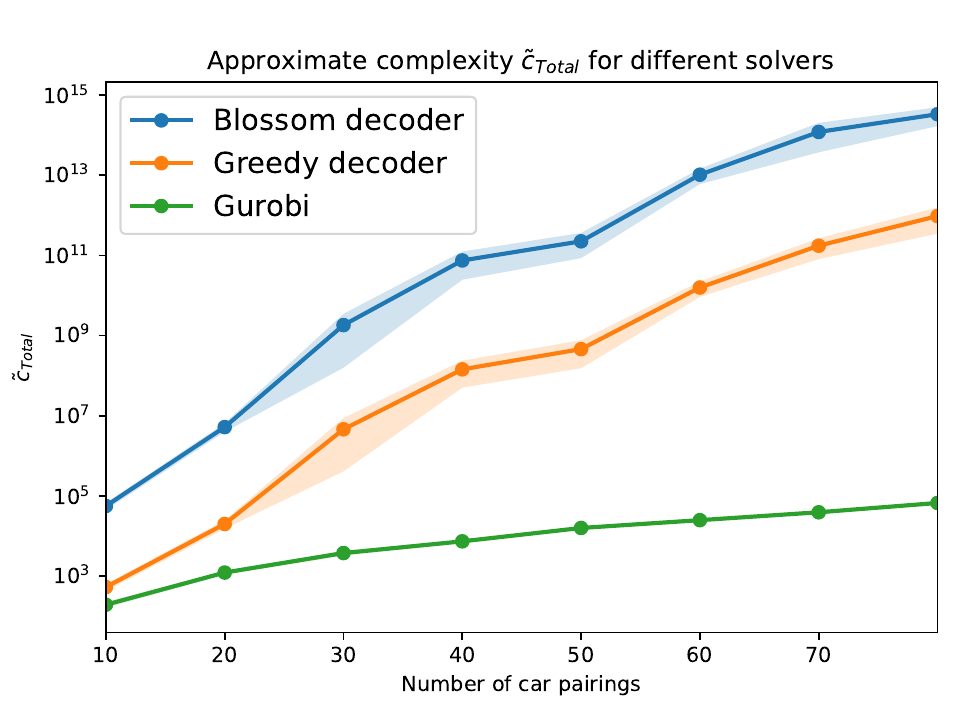}
    \caption{Complexity benchmarking for DQI with both our implementation of the greedy decoder and the hypothetical blossom decoder against Gurobi. As explained in Section~\ref{sec_Approx}, we compute $\tilde{c}_{Total}$ as the product of expected number of times we need to run DQI and the individual runtime of each DQI iteration $c_{DQI}$. For the greedy decoder, we take $c_{DQI}$ to be the leading order contribution to the number of quantum gates in the quantum circuit implementation as explained in Section~\ref{sec_BPSP}. For the hypothetical blossom decoder, we take $c_{DQI}$ to be $n^4$, the leading order contribution in the original blossom algorithm. For Gurobi, we take $\tilde{c}_{Total}$ to be the internal work number for the Gurobi run, multiplied by a factor of $10^6$ to facilitate the visual comparison of the results. The light areas correspond to a third of the standard deviation of the distribution of the obtained results. }
    \label{Fig_complexity}
\end{figure}

Additionally, we want to benchmark this complexity/runtime estimate against the time it takes a state of the art classical solver to find an optimal solution. To do so, we solve each of the BPSP instances with Gurobi (i.e.~we let Gurobi find a single optimal solution) and export the units of ``work" it takes Gurobi to do so. These units of work are used internally by Gurobi to quantify the amount of resources needed to solve the problem. As these are not directly comparable to the number of quantum gates needed (our benchmark for DQI), we are primarily interested in the scaling of these two resource estimations as the problem size grows. To facilitate the comparison of Gurobi and DQI, we multiply the amount of work by a factor of $10^6$, i.e., we set $\tilde{c}_{Total}:=10^6 \cdot \text{work}$  in the case of Gurobi. The results of this benchmarking are visualized in figure~\ref{Fig_complexity}.

We see that the greedy decoder indeed manages to beat the minimum-distance decoder on all tested instances of the BPSP. Unfortunately, DQI with the greedy decoder still performs substantially worse than the state of the art classical solver Gurobi on the BPSP instances in question. While it is at least partially successful to decode errors of Hamming weight well beyond the code distance $d^\bot$ while preparing the DQI state, this is not sufficient to compete with the runtime growth of the classical solver. This might be different on larger BPSP instances, where classical methods become less efficient. However, this would require more extensive testing, respectively, a even less computationally intensive approximate benchmarking scheme. While it is possible to decode well beyond the code distance of a LDPC code in order to improve the DQI state, it still seems very difficult to obtain good results with DQI on problems where the corresponding LDPC code has such a low code distance.

\section{Outlook}

In the present article, we develop a method for obtaining instance-specific benchmarks for DQI in the regime of imperfect decoding. The exact method of our benchmarking scheme involves evaluating the decoder on all bit strings up to a certain Hamming weight which is the computational bottleneck of our benchmarking method.
This can be circumvented by employing an approximate benchmarking scheme, relying on a Monte-Carlo estimation of the decoder error rate.
Empirical results show that the estimated DQI success probabilities obtained in this way serve as a slightly conservative approximation to the actual success probabilities.

We apply our benchmarking scheme to the Binary Paint Shop Problem (BPSP) which admits a natural max-XORSAT formulation.
We examine its code distance, which turns out to be approximately constant, independently of the problem size.
Nevertheless, optimal polynomial degrees for DQI turn out to grow roughly linearly with problem size, necessitating analysis of the imperfect decoding regime.
We employ a greedy algorithm for the decoding problem, which can be reduced to a (minimum) perfect matching problem on a graph induced by the constraint structure of the BPSP.
We provide a simple implementation of the greedy decoder as a quantum circuit consisting entirely of CNOT and Toffoli gates, utilizing that a large portion of the data needed for decoding can be efficiently precomputed classically.

We compare the approximate DQI results obtained in this way with hypothetical results which would result from using a decoder which solves the minimum perfect matching problem exactly.
We find that the greedy-decoder version of DQI consistently outperforms DQI with a more exact, but slower, decoder.
This points to an important tradeoff between decoder performance and speed. As DQI (like most quantum optimization algorithms) is a probabilistic algorithm, success probabilities can be boosted by repeated runs of the algorithm. This gives a faster but less accurate algorithm the opportunity to outperform its slower but more accurate counterpart.

However, a comparison of the scaling behavior for both versions of DQI with Gurobi indicates that we cannot expect a quantum advantage through DQI on the BPSP.
We suspect that the reason for this ultimately lies in the constant code distance obtained from the BPSP which fundamentally limits decoder performance.

In order to move the search for practical applications of the DQI forward, we believe that several puzzle pieces are missing.

First, it is difficult to quantify the suitability of an LDPC code corresponding to a given problem for the DQI algorithm. While the notion of code distance captures the threshold up to which perfect decoding is possible, it does not allow to predict the behavior of the decoding algorithm in the regime of imperfect decoding which in practice seems critical for the performance of DQI.

Another interesting direction for further research is the study of more elaborate encodings for a wide range of problems, with special attention to the properties of the resulting code.

Finally, the field of decoding merits further study from a quantum computing perspective. Realizing elaborate classical decoding algorithms as quantum circuits is no trivial matter.
Thinking further, purely quantum decoders, as opposed to quantum implementations of classical decoders could advance the cause of DQI even more.

\bibliographystyle{alphaurl}
\bibliography{sample}

\end{document}